\documentclass[manuscript]{aastex}

\newcommand{\lta}{\raisebox{-0.6ex}{$\,\stackrel
{\raisebox{-.2ex}{$\textstyle <$}}{\sim}\,$}}
\newcommand{\gta}{\raisebox{-0.6ex}{$\,\stackrel
{\raisebox{-.2ex}{$\textstyle >$}}{\sim}\,$}}   

\shorttitle{COSMIC}
\shortauthors{McCulloch et al.}

\begin{document}


\title{COSMIC : Microarcsecond resolution with a 30 metre radio telescope}


\author{Peter M. McCulloch, Simon P. Ellingsen}
\affil{School of Mathematics and Physics, University of Tasmania, 
  Private Bag 37, Hobart, TAS 7001, Australia}
\email{Peter.McCulloch@utas.edu.au, Simon.Ellingsen@utas.edu.au}

\author{David L. Jauncey}
\affil{Australia Telescope National Facility, CSIRO, P.O. Box 76, Epping, 
  NSW 1710, Australia}
\email{David.Jauncey@csiro.au} 

\author{Steven J.B. Carter\altaffilmark{1}, Giuseppe Cim\`o}
\affil{School of Mathematics and Physics, University of Tasmania, 
  Private Bag 37, Hobart, TAS 7001, Australia}
\email{Steven.Carter@utas.edu.au,Giuseppe.Cimo@utas.edu.au}

\author{James E.J. Lovell}
\affil{Australia Telescope National Facility, CSIRO, P.O. Box 76, Epping, 
  NSW 1710, Australia}
\email{Jim.Lovell@csiro.au}

\and

\author{Richard G. Dodson\altaffilmark{2}}
\affil{School of Mathematics and Physics, University of Tasmania, 
  Private Bag 37, Hobart, TAS 7001, Australia}
\email{rdodson@vsop.isas.ac.jp}

\altaffiltext{1}{School of Engineering, University of Tasmania, Private Bag 65,
  Hobart, TAS 7001, Australia}
\altaffiltext{2}{ISAS, 3-1-1 Yoshinodai, Sagamihara, Japan}

\begin{abstract}
  
  Interstellar scintillation has been conclusively demonstrated to be
  the principal cause of the intraday variability (IDV) observed in
  the centimetre-wavelength emission of many AGN.  A few sources show
  large amplitude modulation in their flux density on a timescale of
  hours.  However, the majority of IDV sources exhibit variability on
  timescales of a day or more.  On timescales of a year some sources
  have been found to show an annual cycle in the pattern of their
  variability. Such an annual cycle occurs because the relative speeds
  of the Earth and the interstellar medium change as the Earth orbits
  the Sun. To search for these annual variations as well as to follow
  the source evolution, requires a dedicated instrument; the necessary
  amounts of observing time are beyond the capability of the national
  facility instruments. Here we describe the scientific motivation
  for, and present an outline of the COSMIC (Continuous Single dish
  Monitoring of Intraday variability at Ceduna) project which uses the
  University of Tasmania's 30~m diameter radio telescope at Ceduna,
  which has been monitoring the flux density of a number of the
  stronger southern scintillators at 6.65~GHz since March 2003.

\end{abstract}


\section{Introduction} \label{sec:intro}

In the last few years interstellar scintillation (ISS), has been
demonstrated to be the principal mechanism responsible for the
intraday variability (IDV) seen in many compact flat-spectrum radio
sources at centimetre wavelengths. Two types of observations have been
responsible for this: a time delay in the arrival times of the
intensity pattern at widely spaced telescopes, and an annual cycle in
the timescale of the variability.  A time delay of several minutes has
been found for the three most rapidly variable sources
PKS\,B0405$-$385 \citep{JKL+00}, PKS\,B1257$-$326 \citep{BJL+04} and
CLASS\,J1819+3845 \citep{DD02}.  Such measurements can be made only for the
most rapid variables with hour or sub-hour timescales, and are not
possible for the more common intraday and inter-day variables.
         
The speed of the local interstellar medium (ISM), is close to the
speed of the Earth in its orbit around the Sun (30~kms$^{-1}$) and
this leads to an annual cycle in the timescale of variability.  For
part of the year the Earth and the ISM are travelling roughly
parallel, the relative velocity is small and the timescale of
intensity changes is long. Six months later the Earth is moving in the
opposite direction, the relative velocity is high and the variability
timescale is much shorter. Such an annual cycle has been found for the
two long-lived rapid variables PKS\,B1257$-$326 \citep{BJL+03} and
CLASS\,J1819$+$3845 \citep{DD02}, as well as for several of the slower IDV
sources : B0917$+$624 \citep{RWK+01,JM01}, PKS\,B1519$-$273
\citep{JJB+03}, and more recently B0059$+$581 (D. L.  Jauncey et al., in
preparation).  These rapid variables have played a crucial role in
establishing ISS as the principal mechanism responsible for centimetre
wavelength IDV.  Unfortunately, there are very few such rapid
variables.  The MASIV survey \citep{LJB+03} showed that the great
majority of sources possess variability timescales of a day or more,
rather than the hours or less of the rapid scintillators.

Measurement of an annual cycle can be used to probe the structure of
both the ISM and the source (on microarcsecond scales), and has been
dubbed ``Earth orbit synthesis'' \citep{MJ02}.  This technique has
been sucessfully applied to determine the microarcsecond flux density
and polarization structure of PKS\,B0405$-$385 \citep{RKJ02} and the
microarcsecond structure of CLASS\,J1819+3845 \citep{DD03} and
PKS\,B1257$-$326 \citep{BJL+03}.  Reliable measurements of an annual
cycle are straight forward for the rapid inter-hour variables, since
many intensity fluctuations or ``scints'' can be seen in a 12 hour
observing session, such as is readily available on national facility
instruments like the ATCA, the VLA and the WSRT.  However, such
measurements for the more common intraday variables require observing
sessions spanning many days on multiple occasions throughout the year.
These are much more difficult programs for the national facility
instruments to accommodate because of the limited time available and
their oversubscription.
         
There are a large number of small (20-40 metre diameter) stand-alone
radio telescopes throughout the world, which cannot compete in terms
of sensitivity with modern large apertures and interferometers.  These
small telescopes often play important roles in VLBI arrays such as the
European VLBI Network (EVN) and Australian Long Baseline Array (LBA),
but struggle to find high-impact science to undertake when operating
as a single dish.  Small (1-2 metre diameter) optical telescopes faced
a similar problem a decade or so ago, however the advent of
microlensing and high redshift supernovae searches provided new
purposes for this class of instrument.  Clearly the niche for small
radio telescopes requires projects which involve large amounts of
observing time, but only modest sensitivity.  To date, this niche has
been filled by projects such as searches for maser lines
\citep[e.g.][]{EVM+96,SKH+02,VPB+01}, or monitoring of bright pulsars
\citep[e.g.][]{F95,DM97}, masers \citep[e.g.][]{GGV04}, or the total
intensity of AGN \citep[e.g.][]{AAH03}.

Following the initial successes in establishing ISS as the cause of
centimetre IDV through the measurement of annual cycles, it became
clear to us that a wider application of this technique to many
scintillating sources requires a radio telescope dedicated to the task
of precision flux density monitoring, in order to determine the
changes in the variability characteristics of a number of sources over
the course of a year or more.  A sufficient numbers of scintillating
sources with mean flux densities of 1~Jy or more have been found in
the southern sky \citep{KJW+01,BJK+02,B03} to consider using the 30 m
telescope at Ceduna, which is operated by the University of Tasmania.

PKS\,B0405$-$385 was the first of the large amplitude, extremely rapid
variables discovered \citep{KJW+97}.  Here the very rapidity of the
variations, if intrinsic to the source, implied brightness
temperatures of up to 10$^{21}$ K, nine orders of magnitude in excess
of the inverse Compton limit. This forced a major re-appraisal of the
intrinsic variability hypothesis which ultimately led to the first
intensity pattern time delay observations and the unequivocal
demonstration that interstellar scintillation rather than intrinsic
variability was the mechanism responsible for the rapid flux density
variations \citep{JKL+00}.  Subsequent investigations of
PKS\,B0405$-$385 and the other two rapid scintillators
PKS\,B1257$-$326 and CLASS\,J1819+3845, have revealed that the reason for
these fast variations is the presence of nearby ISM scattering
screens, rather than any extreme properties ($T_B > 10^{15}$~K) of the
sources. The rapid variability in each has been shown to be the
product of scattering by ISM plasma at a distance of no more than a
few tens of pc \citep{RKJ02}.  Brightness temperatures of the order of
10$^{12}$~K have been established, closely overlapping those found in
conventional VLBI and space VLBI observations \citep{LJB+03}.
                 
The general argument for nearby screens is straightforward. For a
point source in the weak scattering case at typically 5 - 8 GHz, the
linear size, $r_F$, of the first Fresnel zone is roughly the
characteristic timescale, $t_{char}$, multiplied by the component of
the ISM velocity perpendicular to the line of sight ($V_{ISM}$).  This
in turn is proportional to $\sqrt{\lambda D}$, where $D$ is the screen
distance and $\lambda$ the wavelength.  \citet{N92} shows that in the
case of an extended source ($\theta_S \gta \theta_F$, which is
believed to be the general case for AGN), we have
\begin{equation}
  t_{char} \simeq \frac{r_F}{V_{ISM}} \frac{\theta_S}{\theta_F} = 
  \frac{\theta_S D}{V_{ISM}} \label{eqn:tchar}
\end{equation}
where $\theta_S$ and $\theta_F$ are the angular sizes of the source
and the Fresnel zone respectively.  So the the source angular size is
proportional to $t_{char}$, while the brightness temperature, $T_{B}$,
scales as ${\theta_S}^{-2}$ and hence ${t_{char}}^{-2}$. The rapid
variables with small $t_{char}$ have been shown to be scattered by
nearby screens.  For nearby screens, sources with longer $t_{char}$
have larger angular sizes and hence lower brightness temperatures.

The COSMIC program is specifically directed at the more usual inter-
and intraday variables. These have $t_{char}$ values of days rather
than hours and hence they are more likely to be scattered by more
distant screens and therefore to possess smaller angular sizes.  The
program's principal objective is to closely monitor the 6.65~GHz flux
density of several well documented, large-amplitude southern variables
principally selected from the \citet{KJW+01} survey and subsequently
examined in more detail by \citet{B03}. Since the characteristic
timescales for the sample are of the order a day it is necessary to
measure the flux density throughout the day, and from day-to-day for
up to several weeks, in order to follow the variations in detail.
Observations of each source must then be repeated at regular intervals
over the course of a year, or preferably longer, to establish whether
the signature of an annual cycle is present.
                 
Such a large-scale and intensive observational program has never been
undertaken before, so the COSMIC (COntinuous Single dish Monitoring of
Intraday variability at Ceduna) program was initiated, in part to
establish that such a single dish facility could operate sufficiently
well to achieve this goal.  The value of the COSMIC program extends
beyond confirming that ISS is the cause of centimetre wavelength
variability, since it is clear that scintillators display a broad
spectrum of characteristics in their long-term behaviour.  The
observations are probing a new dimension in variability parameter
space, so serendipity may well play a major role in the outcomes from
the COSMIC program.  For example, PKS\,B0405$-$385 exhibits episodic
behaviour periods consisting of a few months of intense activity
interspersed between several years of relative inactivity
\citep{KJW+97,KJW+01,JKL+00,CEC+04}. In contrast, B0917+628, which had
shown the annual cycle characteristic of ISS for more than a decade
\citep{KWK+99,JM01,RWK+01}, suddenly ceased its scintillation in
September 2000 and subsequently has been remarkably inactive
\citep{FKC+02}.  PKS\,B1519$-$273 has been followed closely at the
ATCA since the discovery of its scintillation in 1994, and continues
to show much the same annual cycle a decade later \citep{JJB+03}, as
is well demonstrated in our current Ceduna data (S.~J.~B. Carter et
al., in prep.).
                 
This episodic behaviour demonstrates that significant changes often
take place in scintillating sources on a timescale of months to
years, and likely also in the turbulent ionized interstellar medium.
The early results from the MASIV VLA scintillation survey
\citep[S.~S.~Shabala et al. in prep.]{LJB+03} reveal that episodic
variables outnumber the persistent variables by two to one.

The Ceduna telescope has not been previously described in the
literature and in \S~\ref{sec:antenna} we give details of its
characteristics.  The COSMIC project is outlined in
\S~\ref{sec:cosmic}.

\section{The Ceduna Radio Telescope} \label{sec:antenna}

\begin{figure}
\epsscale{1.0}
\plotone{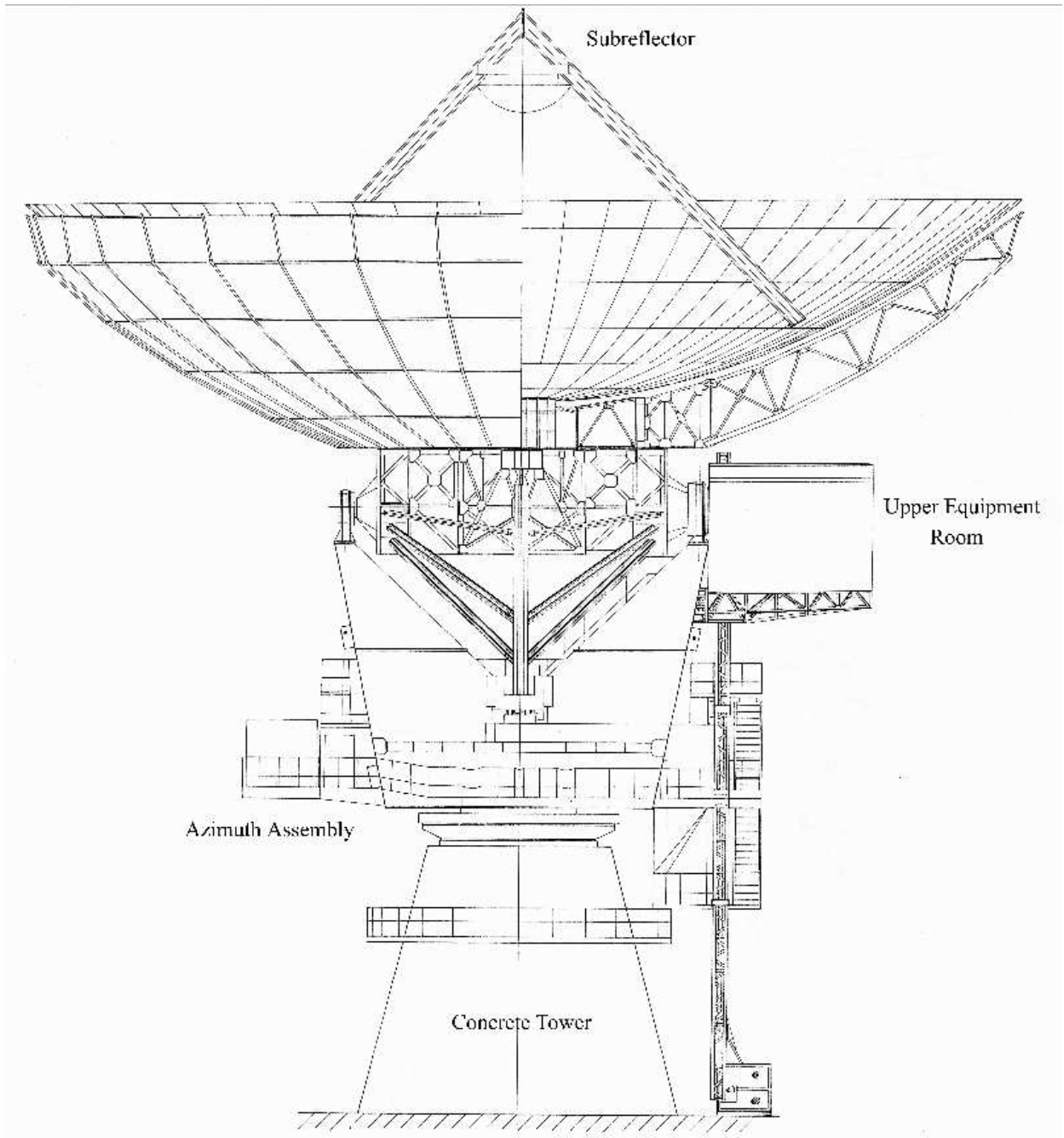}
\caption{Line drawing of the Ceduna 30 metre radio telescope.
 \label{fig:telescope}}
\end{figure}

The University of Tasmania Ceduna radio telescope is a 29.6 metre
diameter parabolic antenna designed and constructed by Mitsubishi in
1969, with an alt-az mount and feeds located at the Nasmyth focus
(Figure~\ref{fig:telescope}).  It is located near the coast of South
Australia, some 850~km west of Adelaide, at longitude
133\degr48\arcmin36.57\arcsec\ east, latitude
-31\degr52\arcmin05.04\arcsec, and an altitude of 161~m above sea
level.  The facility was established by the Overseas
Telecommunications Company (OTC) in 1969, as the Ceduna Satellite
Earth Station and was later taken over by Telstra.  The Earth Station
provided the gateway between Australia and Europe for telephone and
television communication, via the global satellite system set up by
Intelsat that offered sophisticated and low-cost communication
services around the world.  The choice of Ceduna as the location of an
Earth Station was dictated by the limits of the coverage zone of the
Indian Ocean geostationary satellite, the need to be reasonably close
to Australia's populous south eastern region, and the need to be in a
location free from man-made electrical noise.  During 1984, almost
half of Australia's International telecommunication traffic passed
through Ceduna's Earth Station.

In October 1994, improved communication methods (notably fibre optic
links) and a need to rationalise services saw the closure of the
Ceduna Earth Station.  In September 1995 Telstra donated the Ceduna
Satellite Earth Station to the University of Tasmania for use as a
radio astronomy observatory.  At the time of the handover, the antenna
drive and control systems were those originally installed in 1969 and
power was provided by a set of on-site diesel generators.  The
facility's conversion to a radio astronomy observatory was a
significant undertaking and included the connection of mains
electricity, modification of the cable twister assembly, and the
installation of new antenna drive motors and controllers, angle
encoding equipment, feeds and a range of specialised radio astronomy
equipment.  The new electric drive motors provide slew rates of up to
40~degrees per minute about both axes, with an elevation limit of
1\degr.  Angle encoding is highly repeatable and currently an RMS
pointing accuracy of about 20\arcsec\ is achieved.  The COSMIC project
is providing a large amount of additional pointing information which
is expected to result in an improvement of a factor of two or more in
the pointing accuracy.  The antenna surface has been surveyed and
adjusted to an RMS accuracy of about 0.8~mm.

During its operation by the OTC/Telstra, the Ceduna antenna provided
simultaneous up and down-links to the Intelsat satellites using
frequencies in the 4-6 GHz range (transmitting in the range
5.925-6.425~GHz and receiving in the range 3.7-4.2~GHz).  The
high-powered transmission amplifiers originally used with the antenna
were oil-cooled and could not be tilted.  This was the reason for the
choice of Nasmyth optics, since the focus position of the antenna lies
along the elevation axis, allowing the transmitters to be housed in
the ``upper equipment room'' (see Figure~\ref{fig:telescope}).  This
design feature impacts upon the range of frequencies at which the
antenna can operate.  The dimensions of the hole in the elevation axis
limits the diameter of the final section of waveguide feed, resulting
in a lower frequency limit at Ceduna of 2~GHz.  Corrugations in the
large waveguide sections close to the tertiary mirror affect the
propagation at higher frequencies, resulting in 22 GHz being the upper
frequency achievable.  We have designed and built a range of new
terminating sections for the Nasmyth waveguide feed, which operate
in the frequency bands of 2.2, 4.8, 6.7, 8.4, 12.2 and 22 GHz.
For reliability and simplicity the dual circular polarization
receivers at each frequency are uncooled and have a bandwidth of
approximately 500 MHz (except the 22 GHz system).  The frequency range
and system equivalent flux density for each system is summarised in
Table~\ref{tab:receivers}.

\begin{figure}
\epsscale{1.0}
\includegraphics[angle=270]{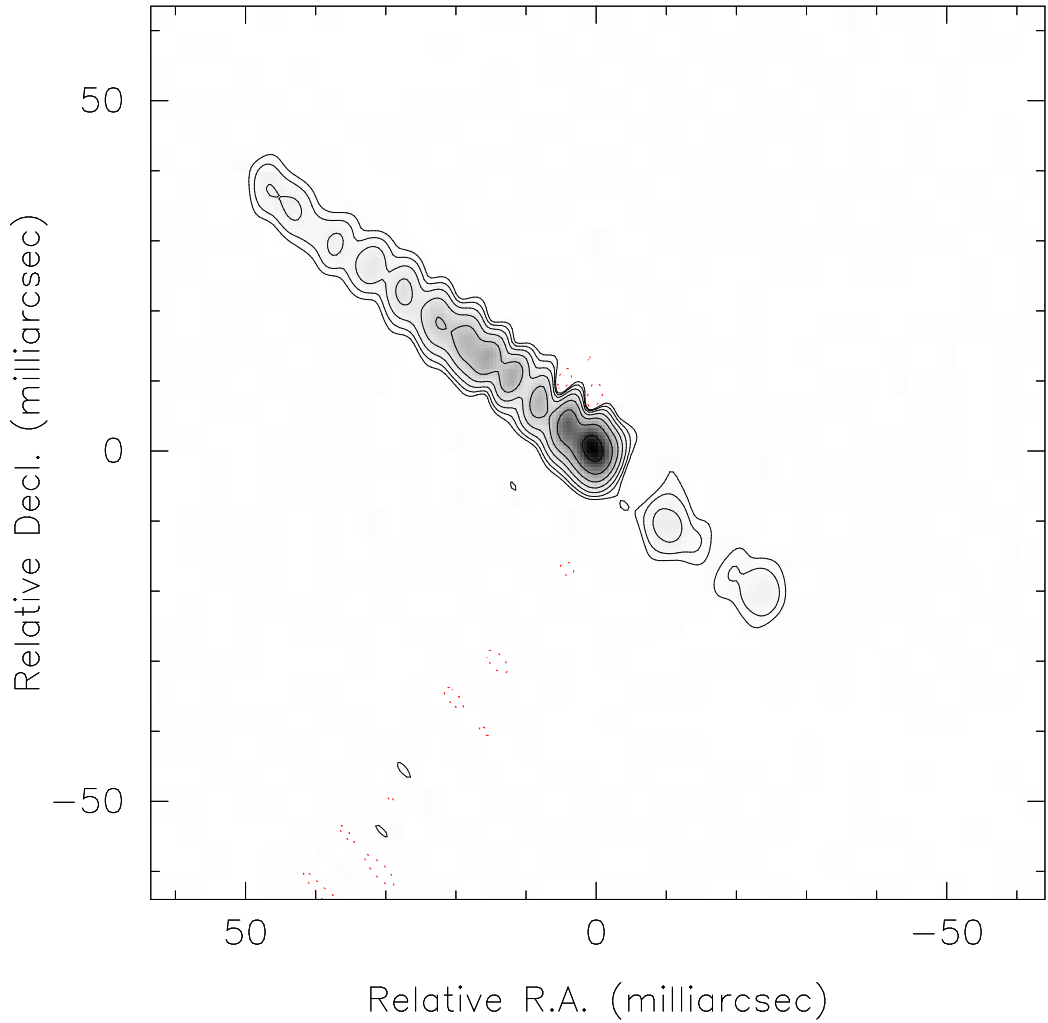}
\caption{The nucleus of the radio galaxy Centaurus A at 6.6~GHz, imaged 
  with the LBA, including Ceduna.  The LBA image clearly shows the
  counterjet and numerous components in the jet.  The contour levels
  are at -1, 1, 2, 4, 8, 16, 32, 64 x 7~mJy~beam$^{-1}$, with negative
  contours indicated by a dashed line.  The peak flux in the image is
  541~mJy~beam$^{-1}$.\label{fig:cena}}.
\end{figure}

The Ceduna 30m antenna conversion project was largely motivated by the
need to extend the Long Baseline Array (LBA)\footnote{The LBA is
  operated as a National Facility by the Australia Telescope National
  Facility and the University of Tasmania.}.  The LBA consists of the
Australia Telescope National Facility, Parkes, Mopra and ATCA
telescopes, the University of Tasmania Hobart and Ceduna telescopes,
with regular participation of the NASA Tidbinbilla telescopes and the
Hartebeesthoek telescope in South Africa.  The Ceduna antenna first
participated in LBA observations in late 1997.  It has significantly
enhanced the imaging capability of the array, by increasing the number
of baselines, and more importantly providing much needed east-west
baselines.  With the inclusion of Ceduna, the longest north-south
baseline (1400 km from the ATCA to Hobart) is comparable to the
longest east-west baseline (1500 km from the ATCA to Ceduna).
Figure~\ref{fig:cena} shows a 6.6~GHz LBA image of the nucleus of the
radio galaxy Centauraus A, in which the counterjet is clearly seen, as
are a number of components along the jet (S.~J.~Tingay et al.  in
prep.).  This image is of significantly higher quality than comparable
images of Centaurus A made with the pre-Ceduna LBA \citep[see for
example Figure~1 of][]{TJR+98}.  The Ceduna antenna is currently
equipped with LBA-DAS/S2 tape-based \citep{WRD95} and Mets\"ahovi
disk-based VLBI recording systems \citep{DTW+04}.

\section{The COSMIC project} \label{sec:cosmic}

The motivation for the COSMIC project is outlined above
(\S~\ref{sec:intro}).  The observations are made using an uncooled
receiver with two orthogonal circular polarizations operating over the
frequency range 6.4-6.9~GHz.  This is in the regime of weak scattering
for most AGN \citep{W98,W01}, but close to the transition frequency
between weak and strong scattering, where the variability amplitude is
large.

To measure the flux density of a source we undertake a series of four
scans, one pair increasing and then decreasing in right ascension
(constant declination), one pair increasing and then decreasing in
declination (constant right ascension).  Hereafter we refer to a
series of four such scans as a scan group.  All scans are centred on
the source position, have a total length of 45\arcmin\ on the sky, and
the telescope is driven at a rate of 3 degrees per min.  Prior to each
scan the strength of the receiver noise diode is measured and compared
to a 1 dB step to determine the system equivalent flux density.
Calibration, interference excision, resampling the intensity data as a
function of position on the sky and slewing of the antenna takes on
average a total of 45 seconds, and results in a single scan taking
approximately 1 minute.  The observing software fits a Gaussian
profile with a quadratic baseline (although for the vast majority of
scans a linear baseline is sufficient) to the intensity data and uses
this fit to measure any pointing offset.  If the increasing and
decreasing scans in the same coordinate yield the same offset then the
source position in that coordinate is updated for subsequent scans.
During the observations the data are sampled at 847 Hz.  At the
completion of each scan an automated procedure is undertaken to remove
the occasional outlying point which differs significantly from the
surrounding samples, and which is caused by interference or occasional
miss-reading of the analogue to digital converters.  After this
process the data are resampled as a function of telescope position
with 187 sample points across a 45\arcmin\ scan.  These data are saved
in FITS format using a local implementation of the single dish
conventions \citep{G00} developed by Chris Phillips.

Post-observation analysis of the data are undertaken using a series of
purpose-written programs in the {\it Matlab} environment.  This
software uses non-linear least squares minimization routines to fit a
Gaussian profile plus a quadratic baseline to each scan (in the same
manner as the observing system).  Practical considerations mean that
the data processing has to be largely automatic.  For this reason one
of the most important aspects of the data processing is the system
quality control routine which ensures that aberrant measurements do
not contaminate the final light-curves.  The first aspect of the the
quality control process is to reject scans which are not well fitted
by the non-linear least squares routine.  The RMS in the residuals
(after subtracting the fitted model to the data), are computed, and
scans with an RMS in excess of a specific (empirically set) tolerance
are rejected.  The second quality control measure is implemented
through consistency checks.  The position of the peak and the FWHM of
the fitted Gaussian profile should be the same for both polarizations,
a significant difference indicating a problem with one or both scans.
Similarly the amplitude, FWHM and position of the Gaussian peak should
be similar for the scans made in increasing and decreasing directions.
Failure of either of the quality control steps leads to rejection of
the entire scan group as it is not possible to confidently measure and
correct for pointing offsets with an incomplete scan group (it is
possible to operate with only a single polarization, but this reduces
the sensitivity by a factor of $\sqrt{2}$).  To ensure an overall high
quality of the data for further analysis (in an unbiased fashion), the
various quality control checks are tightened until approximately 60\%
of scan groups remain.

Scan groups that pass quality control have the data for the two
polarizations scaled to convert their intensity into units of Jy, and
then averaged.  Then the increasing and decreasing scans in right
ascension and in declination are averaged.  The beam pattern of the
Ceduna antenna at 6.65~GHz is well approximated by a circular Gaussian
with a FWHM of 5.8\arcmin.  When there is an offset between the
predicted and actual source position, the flux density measured is
reduced.  However, the right ascension scan measures the declination
component of any offset, and vice versa, allowing scaling corrections
to be calculated.  After application of the calculated pointing offset
scaling factor we have two independent flux density measurements from
each scan group, which translates into approximately 400 independent
flux density measurements (across all sources) each day.  Observations
of a number of calibrator sources shows that the gain of the antenna
varies slightly with elevation.  The flux density measurements are
scaled using the gain-elevation correction $S$ from
Equation~\ref{eqn:gain}.  The gain/elevation dependence was determined
from observations at elevations greater than 10\degr\ (the lower
elevation cutoff chosen for COSMIC) of a number of strong calibration
sources.  The data from each source were normalised to an intensity of
1 at an elevation of 50\degr, and a second order polynomial was fitted
to the normalized data.  The observed gain variations are less than
5\%.
\begin{equation}
  S = 3.722 \times 10^{-5}\ \mbox{el}^2 - 3.481 \times 10^{-3}\ \mbox{el} + 
  1.081 \label{eqn:gain}
\end{equation}

We have examined the RMS for the light curves of three calibrator
sources (PKS\, B1934$-$638, PKS\,B0945+076 and PKS\,B1921$-$293).  We expect
the measured RMS in these calibrators to consist of a constant term,
and term which is some fraction of the source flux density.  These
terms are independent and add in quadrature.  Fitting the observed RMS
of the calibrator sources with this type of function gives a constant
error term of 33~mJy and a fractional term of 1\% of the source flux
density (correlation coefficient 0.997).  So for a scintillating
source like PMN\,J1326$-$5256 which has a flux density of
approximately 1.5~Jy at 6.65~GHz, we expect a measurement error of
approximately 36~mJy or 2.4\% of the mean flux density.  If the data
are smoothed using a 5-point running mean (which corresponds to a
timescale of approximately 20~minutes) the measured RMS for the
calibrators are approximately a factor of 2 lower, as is expected.
The RMS confusion error for a 30~m telescope at 6.65~GHz is
approximately 20~mJy, which is smaller than the 33~mJy noise error
noted above, so confusion is expected to make a small contribution to
the measured scan noise and is not a limiting factor in the current
system.

\begin{figure}
\includegraphics[angle=270,scale=0.7]{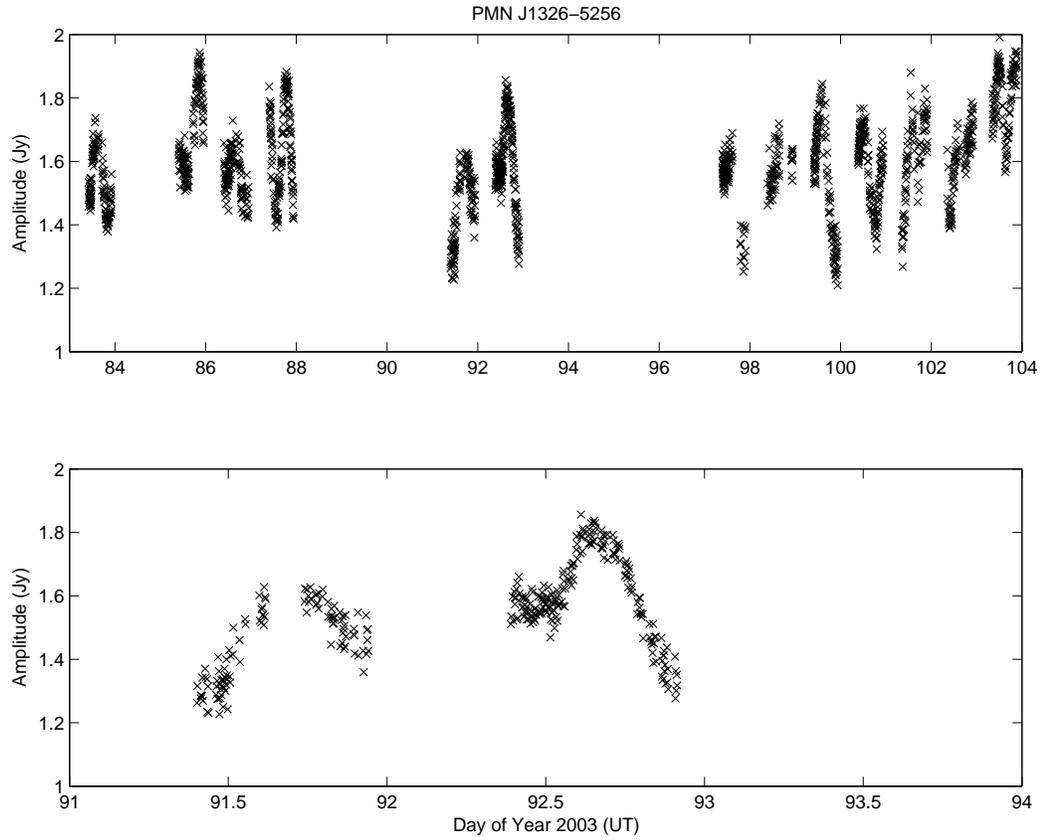}
\caption{(top)Intraday variability in PMN\,J1326$-$5256 observed over a 21 
  day period in 2003 as part of the COSMIC project. (bottom)An
  expanded view of three days observations demonstrates that the system
  is able to track changes in the total intensity of 100~mJy and less.
  \label{fig:idv}}
\end{figure}

\begin{figure}
\includegraphics[angle=270,scale=0.7]{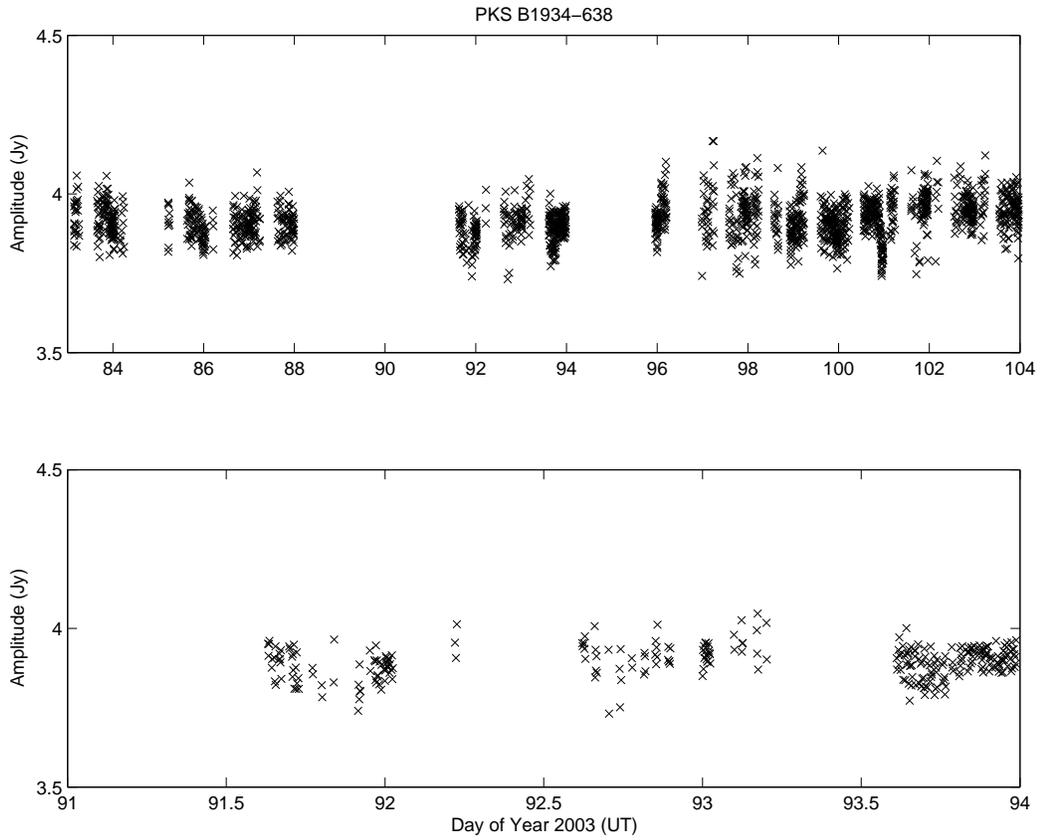}
\caption{(top)COSMIC observations of PKS\,B1934$-$638 observed over the same 
  21 day period in 2003 as Fig.~\ref{fig:idv}. The RMS of the
  individual flux density measurements over this period is 59~mJy.
  (bottom)An expanded view covering the same three period as in
  Fig.~\ref{fig:idv}. The RMS of the individual flux density
  measurements over this period is 49~mJy. \label{fig:cal}}
\end{figure}

The COSMIC project commenced in March 2003 and is monitoring a number
of IDV sources.  The source names, positions and other details are
given in Table~\ref{tab:cosmic}.  The sources are divided into two
separate groups, each of which is observed for alternating periods of
10--14 days.  The length of the period is choosen to allow multiple
scintles to be observed and hence enable the determination of the
characteristic timescale of the variations.  We observe a small number
of sources in each group so as to obtain good sampling for the
light-curves.  The two groupings are: those sources which pass north
of the zenith at Ceduna (AO\,B0235+164, PKS\,B1519$-$273,
PKS\,B1622$-$253), and those that are south of the zenith
(PKS\,B1144$-$379, PMN\,J1326$-$5256).  For each group of sources we
also observe a calibrator, PKS\,B0945+076 (3C227) and PKS\,B1934$-$638
for the northern and southern sources respectively.  All the
scintillating sources have mean flux densities at 6.65~GHz that
typically lie in the range 1--4~Jy and so the measurement error in our
time series can be assumed to be $\lta$50~mJy.  Figure~\ref{fig:idv}
shows 24 days of observations of PMN\,J1326$-$5256 made with the
Ceduna telescope, in which the intraday variability is clearly
demonstrated.  Figure~\ref{fig:cal} shows the calibrator
PKS\,B1934$-$638 observed over the same period, demonstrating that the
variability seen in PMN\,J1326$-$5256 is real and not due to
observational errors.

One unusual aspect of the COSMIC project is that it operates almost
entirely remotely.  The Ceduna antenna is approximately 1700 km from
the University of Tasmania's main campus at Hobart, and the
observatory is only staffed for a few weeks each year during VLBI
observations.  The operation of the telescope is checked daily by a
member of the team based in Hobart, and the data for the previous day
transferred to Hobart for analysis.  A perl script is used to monitor
the operation of the telescope and observing software and to
automatically recover from common problems or issue an alert when it
is not successful.  This has proved to be remarkably successful, with
human intervention seldom required more often than once every 2 weeks.
During the first year of the program (with the exception of VLBI
observing periods) only once were observations interrupted for more
than 12 hours, due to the failure of the noise diode.

\section{Conclusions}

The COSMIC project is a unique experiment to provide quasi-continuous
flux density monitoring that enables the exploration of the
variability of sources on timescales ranging from hours to years. This
will enable us to determine both the microarcsecond structure of the
sources being monitored, and also to better understand the turbulent
local interstellar medium. The sources presently being monitored
predominantly exhibit timescales of days or longer and so cannot be
intensely monitored with national facility instruments such as the
ATCA, VLA or WSRT.

We have demonstrated that we can measure changes in source intensity
with an accuracy and reliability of a few percent for sources stronger
than 1 Jy.  These results will be detailed in upcoming papers which
will also describe the methods used to determine the timescale and
other characteristics of the variations.

The COSMIC project demonstrates that it is possible to undertake
accurate long-term flux density monitoring of appropriate sources with
a 30~m radio telescope.  There are many other scintillating sources
that could be monitored by other similar telescopes.  We also expect
that monitoring of sources in the COSMIC sample at other frequencies,
or by telescopes at different longitudes, will significantly enhance
our ability to interpret and understand the variability in these
sources.

\section{Acknowledgements}
Funding for conversion of the Ceduna antenna for radio astronomy was
provided by Telstra, the Australian Research Council and through the
Major National Research Facilities programme.  Financial support for
the COSMIC project is provided by the Australian Research Council.
The ATNF is funded by the Commonwealth Government for operation as a
National Facility by CSIRO.  We thank Bev Bedson for her vital
contribution to the operation of the Ceduna observatory and the COSMIC
project.  We thank S. Tingay for making available the 6.6~GHz Cen A
image prior to publication.  This research has made use of NASA's
Astrophysics Data System Abstract Service.

\clearpage
\begin{deluxetable}{ccc}
\tablecaption{Characteristics of the receiver systems available at the 
  Ceduna radio telescope}
\tablehead{
\colhead{Band}        & \colhead{Frequency} & \colhead{System Equivalent} \\
\colhead{Designation} & \colhead{Range}     & \colhead{Flux Density}\\
                      & \colhead{(MHz)}     & \colhead{(Jy)}
}
\startdata   \label{tab:receivers}
S        & $2000-2350$   & 400 \\
C        & $4800-5300$   & 450 \\
Methanol & $6400-6900$   & 550 \\
X        & $8200-8700$   & 600 \\
K$_{u}$  & $12100-12600$ & 750 \\
K        & $18000-25000$ & 2500\tablenotemark{a} \\
\enddata
\tablenotetext{a}{The K-band system has only been tested at 22.2~GHz 
  and how the SEFD changes over the frequency range is not yet known}
\end{deluxetable}

\clearpage
\begin{deluxetable}{lccccrr}
\tablecaption{Sources observed as part of the COSMIC program.}
\tablehead{
\colhead{Source} & \colhead{Right}     & \colhead{Declination} & 
  \colhead{Source} & \colhead{Flux Density} & \colhead{Ecliptic}  & 
  \colhead{Ecliptic} \\
\colhead{Name}   & \colhead{Ascension} &                       & 
  \colhead{Type}   & \colhead{Range}        & \colhead{Longitude} & 
  \colhead{Latitude} \\
                 & \colhead{(J2000)}   & \colhead{(J2000)}     & 
                   & \colhead{(Jy)}         & \colhead{(degrees)} & \colhead{(degrees)} \\
}
\startdata   \label{tab:cosmic}
AO\,B0235+164    & 02:38:38.9 & +16:36:59 & IDV        & 1.5 -- 2.0 & 42.46 &   1.09 \\
PKS\,B0945+076 (3C227) & 09:47:46.4 & +07:25:12 & Calibrator & 1.9        &       & \\
PKS\,B1144$-$379   & 11:47:01.4 & $-$38:12:11 & IDV        & 1.5 -- 2.5 & 194.66 & -35.81 \\ 
PMN\,J1326$-$5256 & 13:26:49.2 & $-$52:56:44 & IDV        & 1.2 -- 2.0 & 222.38 & -39.36 \\
PKS\,B1519$-$273   & 15:22:37.7 & $-$27:30:11 & IDV        & 1.9 -- 3.0 & 235.33 &  -8.67 \\
PKS\,B1622$-$253   & 16:25:46.9 & $-$25:27:39 & IDV        & 2.4 -- 3.2 & 248.80 &  -3.74 \\
PKS\,B1934$-$638   & 19:39:25.0 & $-$63:42:46 & Calibrator & 3.9        &        &        \\
\enddata
\end{deluxetable}



\clearpage

\clearpage

\clearpage

\end{document}